\documentclass[preprint,12pt]{elsarticle}

\usepackage{amssymb}
\usepackage{multirow}

\usepackage[boxed]{algorithm2e}

\journal{IJHPCA}

\def\RR{\overrightarrow{R}}
\def\TT{\overrightarrow{T}}

\begin{document}

\begin{frontmatter}



  \title{Accelerating Radio Astronomy Cross-Correlation with Graphics
    Processing Units}


\author{M.~A.~Clark}
\address{Harvard-Smithsonian Center for Astrophysics, \\60 Garden St, Cambridge,  MA 02138, USA}

\author{P.~C.~La~Plante}
\address{Loyola University Maryland, Baltimore, MD 21210, USA}

\author{L.~J.~Greenhill}
\address{Harvard-Smithsonian Center for Astrophysics, \\60 Garden St, Cambridge,  MA 02138, USA}
\begin{abstract}
  We present a highly parallel implementation of the cross-correlation
  of time-series data using graphics processing units (GPUs), which is
  scalable to hundreds of independent inputs and suitable for the
  processing of signals from ``Large-\(N\)'' arrays of many radio
  antennas.  The computational part of the algorithm, the X-engine, is
  implementated efficiently on Nvidia's Fermi architecture, sustaining
  up to 79\% of the peak single precision floating-point throughput.
  We compare performance obtained for hardware- and software-managed
  caches, observing significantly better performance for the latter.
  The high performance reported involves use of a multi-level data
  tiling strategy in memory and use of a pipelined algorithm with
  simultaneous computation and transfer of data from host to device
  memory.  The speed of code development, flexibility, and low cost of
  the GPU implementations compared to ASIC and FPGA implementations
  have the potential to greatly shorten the cycle of correlator
  development and deployment, for cases where some power consumption
  penalty can be tolerated.
\end{abstract}


\end{frontmatter}


\section{Introduction}
\label{sec:introduction}

We apply graphics processing units (GPUs) to the problem of signal
processing for radio astronomy.  While not a classic high performance
computing (HPC) application, there are now many radio astronomy
applications that require in excess of O(100) tera-floating-point
operations per second (TFLOPS) sustained performance.  The Square
Kilometer Array (SKA) due to be built circa 2020, will raise the
processing needs into the exascale regime~\cite{SKA, cornwell-exaflop}.


A typical processing pipeline consists of: digitization of the raw
voltage time-series from individual antennas; cross-correlatation;
instrument electronics calibration and Fourier imaging reconstruction
of the sky.  The extreme computational cost lies predominantly in the
cross-correlation stage; this requires that the signal from every
antenna is correlated with every other, and scales quadratically with
the number of antenna.\footnote{Alternative approaches that scale as
  \(N \log N\) have been proposed~\cite{PhysRevD.82.103501}, but
  cross-correlation, thus far the mainstay technique, enables greater
  generality in data calibration and range of scientific application.}
Science applications that demand high dynamic range and sensitivity in
imaging drive interest in arrays of hundreds to tens of thousands of
collectors (e.g., antennas).  This raises the processing needs from
being modest and easily manageable, into the HPC domain, e.g., the
Murchison Widefield Array, currently in prototype stage in Australia
will require O(10-100) TFLOPS sustained for cross-correlation of
configurations from 128 to 512 collectors~\cite{Lonsdale:2009cb}.



The application of GPUs to cross-correlation is not uncharted
territory: there have been several works investigating GPUs use for
this very purpose~\cite{Harris:2008, Wayth:2009,
  Nieuwpoort09usingmany-core}.  While these demonstrated that GPUs are
suitable for this task, in all cases only 10--30\% of the GPU's peak
performance was obtained and the problem was described as being {\it
  bandwidth bound}.  This work presents an approach to
cross-correlation that is catered to the deep memory hierarchy of
Nvidia's Fermi GPU.  We compare the performance obtainable using both
hardware- and software-managed caches, the latter of which is more
familiar to GPU programmers.  We find in favor of the software-managed
cache, achieving up to 79\% of peak performance, equating to
performance in excess of 1 TFLOPS on the GeForce GTX 480.  More
significantly, our approach is scalable to future architectures which
will likely feature a greater disparity between compute throughput and
memory bandwidth.


This paper is split up as follows; in \S\ref{sec:cross-correlation} we
describe the cross-correlation process, specifically the FX algorithm.
An overview of GPUs and previous attempts at using GPUs for this
problem is given in \S\ref{sec:gpus}.  We describe our kernels in
\S\ref{sec:implementation}, where we contrast the hardware- and
software-managed cache implementations.  In \S\ref{sec:pcie} we
consider performance of the integrated system where we include the
overhead of PCIe transfers.  We discuss the implications of our
results in \S\ref{sec:discuss} before concluding in
\S\ref{sec:conclusions}.



\section{Cross-Correlation}
\label{sec:cross-correlation}

\subsection{XF and FX Correlators}
\label{sec:xf-fx}
The Raleigh criterion \(\theta = \frac{\lambda}{d}\) states the
angular resolution \(\theta\) achievable through direct observation
from focusing optics of diameter \(d\) observing light of wavelength
\(\lambda\).  At the low-end, at radio frequencies, this would require
optics of improbable diameter.  It is thus necessary to use
interferometry, in particular {\it synthesis imaging}, where a
two-dimensional array of antennas (or {\it stations} of many clustered
antennas) act cooperatively as a single virtual telescope, with the
virtual diameter given by length of the farthest distance between the
antennas (see \cite{tms} for a review of synthesis imaging).  This
allows an extension of the Raleigh criterion to larger physical
scales.

The cross-power spectrum at frequency \(\nu\) observed by the station
pair \(i\) and \(j\) (known as a {\it baseline}) is given by
\begin{equation}
S_{ij}(\nu) = \int_{-\infty}^{\infty} (A_i \star A_j)(\tau) \, e^{-i2\pi\nu\tau} d\tau,
\label{eq:power-spectrum}
\end{equation} 
where \(A_k\) is the signal at station \(k\) and \( (A_i \star
A_j)(\tau)\) is the cross-correlation.  With sufficient numbers of
baselines, a detailed power spectrum representation can be obtained,
from which fundamentally an image of the sky can be obtained in turn
through inverse Fourier transform in the spatial domain.  The above
continuum formulation assumes a continuously sampled signal, whereas
in reality the voltage time-series from each antenna will be digitized
and sampled at a given total bandwidth \(B\).  Typical digitization
precision is between two and twelve bits stored in two's complement
form.  The signal is then divided into \(F\) frequency channels via
Fourier techniques in the time domain, where the number of channels is
determined by the needs and goals of the astronomy application.  The
maximum total bandwidth is determined in accordance with the
Nyquist-Shannon sampling theorem.  The channel width \(W = B/F\) is
determined by the number of samples entering the time domain Fourier
transform.

In radio astronomy, the processor that produces the power spectrum
from an array of stations is known as the {\it correlator}.
Historically, this would essentially compute equation
\ref{eq:power-spectrum} directly, i.e., the cross-correlation followed
by the Fourier transform.  Such a correlator is known as a lag
correlator, or XF correlator (the X signifying cross-correlation, and
the F signifying the Fourier transform).  For a correlator that
processes \(F\) frequency channels from \(N\) stations this scales as
\(O(F N^2)\).  

Note, however, from the cross-correlation analogue of
the convolution theorem
\begin{equation}
\mathcal{F}(A \star B) = (\mathcal{F}A) \times (\mathcal{F}B),
\end{equation}
we can write the power spectrum as
\begin{equation}
S_{ij}(\nu) = X_i(\nu)^* X_j^\dagger(\nu),
\label{eq:XF}
\end{equation}
where \(X_i(\nu) = \int_{-\infty}^{\infty} A_i(\tau)
e^{-i2\pi\nu\tau}\) is the Fourier transform of signal from station
\(i\).  Thus, we can calculate the power spectrum from first Fourier
transforming the digitized signal from each station, and then
cross-multiplying the result with every other station's result at
fixed frequency.  Such a correlator is known as an FX correlator,
where the Fourier transforming component is known as the {\it
  F-engine} and the cross-multiplying component is known as the {\it
  X-engine}.  This represents a more cost efficient approach to
computing the cross-power spectrum since the cost scales as \(O(N \ln
F)\)\footnote{For complex-valued samples there are \(N\) (fast)
  Fourier transforms of length \(F\) performed, but only once every
  \(F\) samples, so the total cost is \(O(N \ln F)\).}  and \(O(N^2)\)
for the F and X component,s respectively.  Since the number of
frequency channels is essentially fixed, in the limit of large \(N\)
the X-engine accounts for the bulk of the computational budget.  We
note an additional step between the F- and X-engines, the {\it corner
  turn} which is a reordering of data necessitated by the station
independence of the F-engine, and the frequency independence of the
X-engine.  While this stage involves no computation, its bandwidth
requirements scale as \(O(BN)\) and can be a significant logisical
challenge in terms of the signal routing required.

If we consider all baselines, the X-engine evaluates the sum of a
series of self outer products,
\begin{eqnarray}
\hat{S}(\nu) & = & \sum_{t=1}^{I} {\bf X}_t(\nu) \otimes {\bf X}_t(\nu),
\label{eq:outer}
\end{eqnarray}
where \(\hat{S}(\nu)\) is the correlation matrix and \({\bf
  X}_t(\nu)\) is the vector of signals from \(N\) stations signal at
time \(t\).  This is evaluated independently for every frequency
\(\nu\) and in order to improve the signal-to-noise ratio the outer
product is integrated in time over \(I\) samples.
 
Typically, each station records both the two orthogonal polarizations
of the input signal, which thus doubles the dimension of the input
vectors, hence quadrupling the cross-correlation cost.  Given that the
matrix \(\hat{S}\) is Hermitian, we need only calculate its lower
triangular (or upper triangular) elements, corresponding to
\(\frac{1}{2}N (N+1)\) correlation pairs (including the
auto-correlation of self pairs along the diagonal).  The rate at which
this computation must take place is determined by the total signal
bandwidth \(B\), given by the product of the number of frequency
channels and the channel width.  Written explicitly in terms of
floating point operations per second (FLOPS), the X-engine's compute
requirements are
\begin{equation}
\mbox{FLOPS} = 8 \times B \times \frac{1}{2} 2N(2N+1),
\label{eq:flops}
\end{equation}
where the factor 8 arises from the complex-valued multiply-accumulate
operation and each of factors of 2 multiplying \(N\) arises from dual
polarization.  Note that at fixed total bandwidth there is no explicit
dependence on the number of frequency channels.


\subsection{Characterizing the X-engine}
\label{sec:character}
\begin{figure}
\begin{center}
\includegraphics[width=12cm]{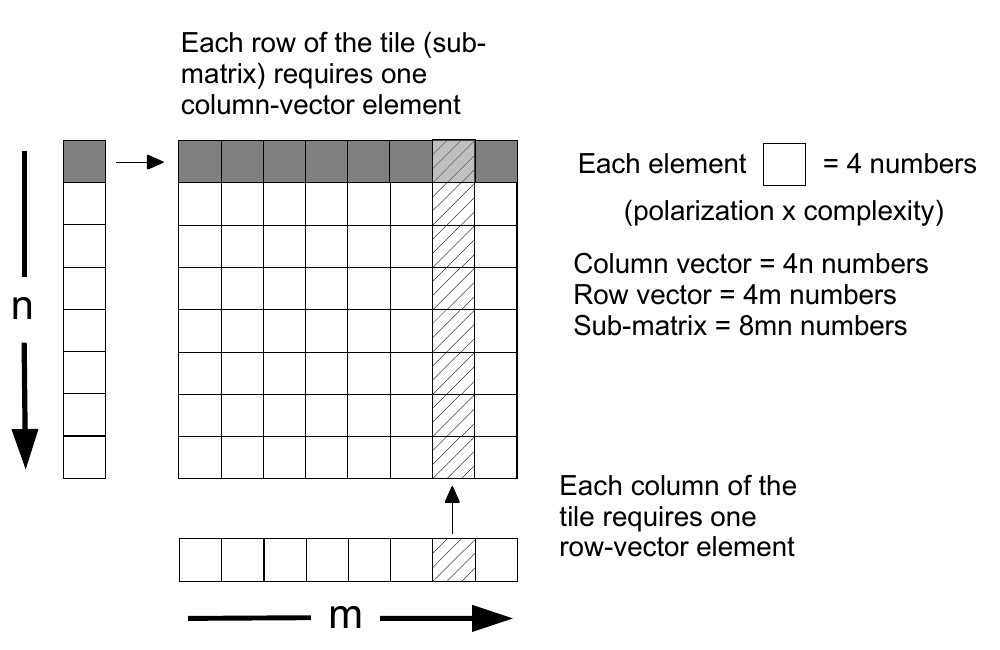}
\end{center}
\caption{An illustration of the tiling strategy together with a breakdown in resources required to achieve this.}
\label{fig:tiling}
\end{figure}

For any computational routine, the critical measure of performance
obtainable is the {\it arithmetic intensity}, or how many floating
point operations are performed per byte of information transferred.
In evaluating Equation \ref{eq:outer} we have to consider both the
input and output bandwidth requirements versus the amount of
computation required.  To calculate each baseline, we must take the
outer-product of two distinct complex-valued vectors of length two and
sum the result to an accumulator.  Assuming 32-bit floating point
data, this requires 32 bytes of input, 32 flops (16 multiply-adds) and
64 bytes for the read and write of the accumulator.  The resulting
arithmetic intensity of \(32/96 = 1/3\) would be disastrous for
performance on current architectures, which typically have a ratio
significantly greater than this to their main-memory space.  This
ratio will only increase with time owing to the increasing energy cost
in moving data relative to operating on
it~\cite{Bergman08exascalecomputing}.  Fortunately, the situation can
be improved by making some simple observations:
\begin{enumerate}
\item If instead of considering a single baseline, we consider a ``tile''
  of baselines of size \(m\times n\),\footnote{Note we adopt
    width\(\times\)height ordering notation for referring to tiles
    and block sizes, not the transpose (matrix notation) with the
    origin lying at the upper-left corner.} there is
  significant data reuse between the baselines and we need only load a
  ``column'' of length \(n\) and ``row'' of length \(m\) to satisfy
  all input memory requirements.
\item The output memory traffic can be supressed by a factor of \(I\)
  if we do not store the accumulated result until we have completed
  calculations for all \(I\) samples.
\end{enumerate}
This tiling strategy is illustrated in Figure \ref{fig:tiling}.  With
these generalizations, the arithmetic intensity is given by
\[
\mbox{Arithmetic Intensity} = \frac{32mnI}{16(m+n)I + 64mn}.
\]
The value of \(I\) is determined by how long the time-series signals
are integrated for; values are application dependent, but for
frequency channel widths of 1 to 100 kHz and accumulation times of 1
to 10 seconds, \(10^4 < I < 10^6\).  At such large \(I\) we see
immediately that the output memory traffic becomes negligible, and we
can make the arithmetic intensity arbitrarily large through increasing
the tile size.  This increases the resources (e.g., registers)
required as \(8mn + 4(m+n)\), which imposes a practical limit upon the
size of the tile.\footnote{This is actually an upper bound, since
  resources can be reclaimed and reused with appropriate
  optimization~\cite{Nieuwpoort09usingmany-core}.}  We demonstrate in
\S\ref{sec:implementation} that this limitation can be overcome on
architectures that feature a multi-level memory hierarchy, through
employing a multi-level tiling strategy, where only at the smallest
tile size (register level) is the matrix ``filled in'', and the other
levels are used to store the input vectors only, thus requiring only
\(4(m+n)\) storage.  Obtaining high performance is thus a balancing
act between maximizing arithmetic intensity and ensuring that
sufficient resources are available.  The computation is similar to
GEMM (dense matrix multiplication) in this regard~\cite{vanLoan}.

\subsection{Hardware Correlators}
\label{sec:fpga}
Application Specific Integrated Circuits (ASICs) and Field
Programmable Gate Arrays (FPGAs) are commonplace computing engines for
large correlators.  These platforms are well suited to the
cross-correlation of radio astronomical time-series data because they
excel at limited precision fixed-point computations and (synchronous)
signal routing, and because they enable fine-grained optimization of
resources.  Typically four to eight bits of precision is sufficient
for both the Fourier transform and cross-multiplication operations,
with a larger number of bits used for the accumululator to prevent overflow.
ASICs offer greater power efficiency since all of the silicon is devoted
specifically to the problem at hand; however, they are expensive to
develop and produce.  FPGAs strike a middle ground between general
purpose commodity processors (e.g., Intel x86) and ASICs, being much
cheaper and easier to apply in development phases than ASICs, because of their
reconfigurability, while being much more power efficient than commodity
processors.  Almost all current radio telescopes under development
plan to use FPGA correlators, e.g., \cite{skamp}, and it is to this
platform that GPUs should be compared.




\section{Graphics Processing Units}
\label{sec:gpus}

\subsection{Fermi Architecture Overview}
\label{sec:fermi}
There are a multitude of overviews of the CUDA architecture and
programming model, we refer the reader to the extensive literature
available, e.g., \cite{cuda}.  Here we focus on significant changes
versus previous CUDA generations and specific architecture features
that are critical for our application.  In the discussion that
follows, we follow standard practice referring to the CPU system
controlling the GPU as the {\it host}, and the GPU as the {\it
  device}.  The program that executes on the device is the {\it
  kernel}.

Fermi provides up to 512 processing cores, arranged in units of 32,
each of which is known as a streaming multiprocessor (SM).  For this
work we used the GeForce GTX 480, which features a Fermi GPU with 480
processing cores for a peak performance of 1345 GFLOPS (counting two
floating point operations from a single fused-multiply-add operation)
connected to 1.5 GiB of off-chip on-card device memory (physically
located on the card containing the GPU).

The programming model is described as being SIMT -- Single Instruction
Multiple Threads, where each group of threads acting cooperatively is
known as a {\it thread block}.  The thread blocks have a 1-, 2- or
3-dimensional cartesian decomposition and themselves reside within a
larger 1- or 2-dimensional cartesian grid.  Each thread block is
assigned to an SM, and depending on the resources available, multiple
thread blocks can be assigned to each SM.  Each thread block is then
subdivided into groups of 32 threads, a {\it warp}, which can be
treated as the SIMD (Single Instruction Multiple Data) width.  While
branching can take place within a warp, such execution is serialized,
and so should be avoided.  The overhead for creating and destroying a
thread is extremely small, with context switching being essentially
free.  Latency is hidden by having many more threads active than there
are cores, so that any threads that are stalled waiting for
instructions to complete can be replaced with threads that are ready
to execute.

\begin{figure}
\begin{center}
\includegraphics[width=10cm]{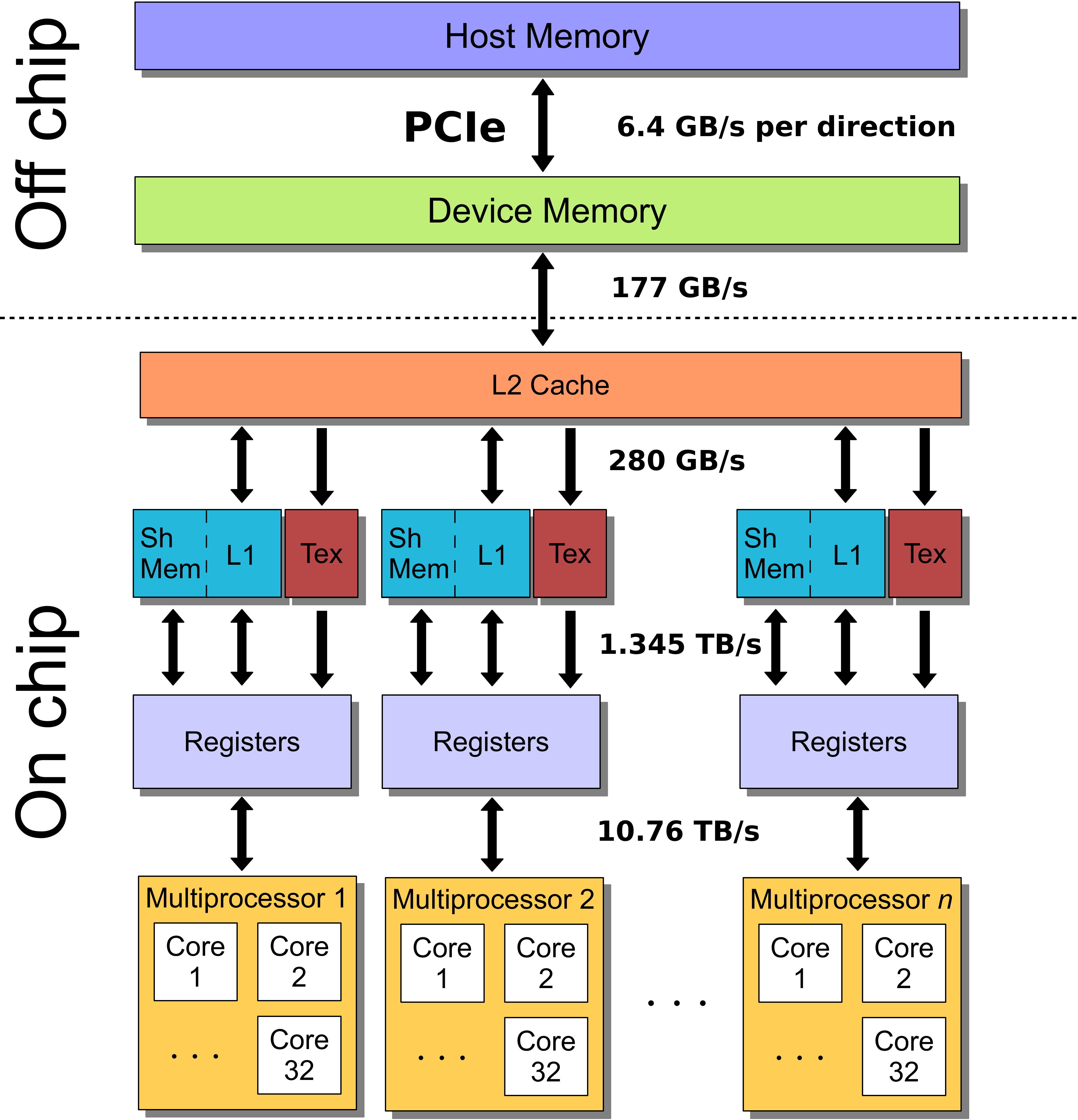}
\end{center}
\caption{A schematic of the memory hierarchy of the Nvidia Fermi
architecture with the peak bandwidth between each layer.  All
numbers quoted correspond to the aggregate bi-directional rate, with
the exception of the PCIe rate which is per direction.  Shared memory (Sh
Mem) and L1 cache (L1) are split between a common 64 KiB cache
(GeForce GTX 480, Tex = texture cache).}
\label{fig:fermi}
\end{figure}

An overview of the Fermi memory hierarchy is provided in Figure
\ref{fig:fermi}.  A significant change from prior generations is the
addition of traditional L1 and L2 caches: the (768 KiB) L2 cache is
shared by all SMs, and there is one L1 cache per SM.  The shared
memory, which is a software-managed cache, and L1 cache are shared
from a common 64 KiB memory, and can be configured in 16/48 KiB or
48/16 KiB partitions, respectively.  When a thread requests data, the
request goes through L1 cache, then L2 cache, and finally off chip to
the device memory. There is a pool of 32768 32-bit registers
available, with a maximum of 63 registers per thread (compared to 128
on pre-Fermi CUDA GPUs).  Fermi is a true load/store architecture with
a unified address space, with the result that threads can no longer
access shared memory operands directly, so all data must first be
copied to the individual registers incurring additional
instructions. Additionally, threads may load data from the texture
unit cache in order to take advantage of array element interpolation
or to avoid polluting the L1 cache.  A feature we make use of is the
free conversion (i.e., without impacting available cycles available
for computation) from 8- or 16-bit integer format data to 32-bit
floating point when the texture cache is used.  Although not relevant
for this work, there exists another small cache, local to each SM, the
constant cache.  This is a read-only cache, and is useful for storing
read-only parameters.

Moving to off-chip device memory, while much slower than on-chip
cache, its total read/write performance of 177 GB/s is much higher
than current typical CPU memory.  Accesses to device memory are high
latency operations, and typically require a high {\it occupancy},
e.g., many concurrent threads, to hide this latency.  Increasing the
number of independent memory transactions per thread can also hide the
latency through instruction level parallelism~\cite{volkov}.  The L1
cache-line size is 128 bytes, and all memory transactions greater than
this are broken down into multiple memory requests.  In order to
achieve close to peak performance, all memory accesses must be {\it
  coalesced}; these are obtained when either a warp, half warp or
quarter warp access consecutive memory blocks of the cache line size.
For sub-128 byte memory access patterns, it can be advantageous to
disable the L1 cache, or to read through the texture cache; here the
L2 cache-line size of 32 bytes determines transaction granularity.
For many applications the main bottleneck is the PCIe bus through
which all transfers to the GPU must take place.  While the PCIe 2.0
x16 specification is quoted at 8 GB/s per direction this does not
account for the 8b/10b encoding used for all PCIe transfers: the
actual peak data rate is 6.4 GB/s per direction.  Communication over
this bus can take place asynchronously, meaning that a kernel can be
executing while memory transfers are taking place.

Communication between threads within a thread block takes place
through the shared memory.  Since the warp execution order cannot be
controlled, race conditions can develop if threads simultaneously
attempt to read and write to a given shared memory address.  To
prevent this, explicit thread synchronization is necessary,
introducing additional latency which can impair performance.  Shared
memory addresses are divided into 32 memory banks in round-robin
fashion, and to enable maximum bandwidth each thread within a warp
must access a unique bank.  The exception to this rule is if threads
are accessing the same address, in which case a broadcast is
supported.  Fermi extends this to support multicast, so that multiple
groups of threads can access multiple addresses simultaneously without
loss of bandwidth.  This functionality is critical to our kernel
implementation as shall be discussed in \S\ref{sec:software-cache}.


\subsection{Overview of Previous Work}

There have been three publications of note regarding the use of GPUs
for cross-correlation.  The first such work \cite{Harris:2008}
implemented the X-engine only, and while the peformance obtained was
reasonable compared to a CPU only implementation, there was little
consideration for how a fully integrated correlator could be
constructed, e.g., considering whether the performance could be
maintained with the PCIe bus constraint.  The X-engine implementation
presented in \cite{Wayth:2009} used similar strategies to those
presented in \cite{Harris:2008} sustaining around 120 GFLOPS at
\(N=32\) on an Nvidia C1060.  This work additionally implemented the
full FX correlator on a single GPU using the CUFFT library (which
comes as part of the CUDA toolkit) for the F-engine.  This approach,
however, suffers from poor scalability at large \(N\) since it cannot
be spread simply across multiple GPUs because of the required corner
turn between the F- and X-engines.  In an exhaustive comparison of
X-engine performance across various multi- and many-core processors
(IBM Cell, IBM BG/P, Intel Core i7, ATi Radeon 4870, Nvidia C1060)
\cite{Nieuwpoort09usingmany-core} found that a fully integrated GPU
X-engine solution would be bound by PCIe bus transfers, restricting
performance to 243 GFLOPS (for \(N = 64\)) on the C1060.  They found
the two IBM platforms achieved a very high peak performance
percentage, with the Cell the most power efficient.

All of the above implementations used memory tiling to improve
arithmetic intensity.  Both \cite{Harris:2008} and \cite{Wayth:2009}
used shared memory tiling to reduce bandwidth pressure to device
memory, whereas \cite{Nieuwpoort09usingmany-core} used register tiling
only, reporting that shared memory tiling was detrimental to
performance.  In this work we show that both tiling methods are
critical to achieve high performance.

\section{X-engine Kernel Implementation}
\label{sec:implementation}

All performance results presented in the section and subsequently were
obtained using a 64-bit Linux workstation, running Ubuntu 10.04, CUDA
3.2 and Nvidia driver version 260.19.26.  The flags passed to the
compiler included ``-X abi=no -m32'' which disables printf support in
CUDA kernels and uses 32-bit pointers, respectively.  We found this
reduced the number of registers required, improving
performance through higher occupancy .

\subsection{Mapping the X-engine to the Fermi Memory Hierarchy}
\label{sec:mapping}

In Table \ref{tab:intensity} we consider the arithmetic intensity
required to achieve peak performance when transferring data from each
of the memory spaces available.  In order to model the X-engine we
assume that the accumulator operands are sourced from registers (and
hence negligible), with the multiply operands sourced from the
respective memory space.  In this naive analysis the required
intensity is given by the ratio of the peak floating point throughput
(1345 GFLOPS) and the peak memory bandwidth achievable from a given
memory space.  This model presumes that all computation and
communication can be overlapped, and neglects the overhead of actual
instruction issue which cannot be overlapped, nor does it include any
required pointer arithmetic, etc.  As we shall see in the following
results, this analysis fails to predict the required arithmetic
intensity for shared memory loads; nevertheless, it serves as a guide
for how to design the kernels.

\begin{table}[h]
\begin{center}
\begin{tabular}{|l|l|c|c|c|}\hline
Level & Memory space & AI  & Square tile size &
Resources \\ \hline
0 & Registers               &  0.125 & 1 & 16 registers\\ \hline
1 & Shared memory / L1  & 1& 8 & 256 bytes \\ \hline
2 & Device memory        & 7.6  & 210 & 6720 bytes\\ \hline
3 & PCIe bus                & 210  & - & \\ \hline
\end{tabular}
\caption{The arithmetic intensity (AI) supported by a given memory
  space and the resulting required square tile size in a given memory
  space to achieve peak performance if the multiply operands are
  serviced from the next slowest memory space (32 bits per real word).
  The last column gives the amount of resources required in each of
  the memory spaces to achieve this tile size (assuming 32-bit
  floating point storage).  }
\label{tab:intensity}
\end{center}
\end{table}

If all memory traffic were to be sourced from device memory, an
8\(\times\)8 register tile size would be required if a simple
one-level tiling strategy is used.  However, with a 63
register-per-thread limit, the maximum tile size achievable is
\(2\times2\).  Thus a multi-level tiling strategy is vital.  In the
second and third columns of Table \ref{tab:intensity} we consider the
required square tile size at level \(i\) given that the memory traffic
originates from level \(i+1\), and the resulting resources required to
achieve this.  At the registers, all data is sourced from the shared
memory / L1 cache requiring a minimum thread granularity of one thread
per baseline, or a register tiling of \(1\times1\).  Since a specific
thread block's assignment is not exposed to the CUDA programmer, in
moving data to shared memory / L1 cache we ignore the L2 cache, and
consider next the device memory.  The resulting \(8\times 8\) tile
size requires 256 bytes of storage, and so is easily achievable with
high occupancy since we have a 48 KiB pool of shared memory to draw
upon.

To overcome the PCIe express bottleneck, the device memory tile size of
\(210 \time 210\) suggests that it will be impossible to feed the X-engine at a
sustained rate for \(N < 210\).  However, note that for considering
the full matrix, the required row is the conjugate of the column,
halving the memory traffic.  We delay further discussion of PCIe
transfers until \S\ref{sec:pcie}.

In terms of data ordering, we have assumed that the corner turn has
been applied, i.e., the input signal ``vectors'' are ordered such that
the station dimension runs faster than the frequency dimension, which
in turn runs faster than the time dimension.  Each signal consists of
two complex-valued polarizations, stored as a float4 (a vector of four
consecutive 32-bit floats).  Thus if adjacent threads are responsible
for loading signals from adjacent stations, full memory bandwidth will
be obtained subject to the constraints described in \S\ref{sec:fermi}.


\subsection{Thread Block Mapping}

Since every station must be correlated with every other, there are
\(\frac{1}{2}N(N+1)\) distinct baselines that must be computed
(including auto-correlations).  This corresponds to the lower
triangular sub-matrix of \({\bf S}(\nu)\) and makes the mapping of
thread index to global-baseline index less straightforward.  Previous
work deployed different strategies for dealing with this mapping:
\cite{Harris:2008} allocated a full 2-dimensional grid of thread
blocks, and if a thread mapped to the matrix strict upper diagonal the
thread exited immediately, doing no work; \cite{Wayth:2009} only
allocated the correct minimum number of thread blocks, where the
mapping was facilitated by a look up table stored in the GPU's
constant cache.  We deemed the former solution to be inelegant, since
it relies on the low overhead of creating and destroying threads, and
the latter approach cannot be scaled to large \(N\) because of the
limited size of constant cache.

\begin{figure}[htb]
\begin{center}
\includegraphics[width=10cm]{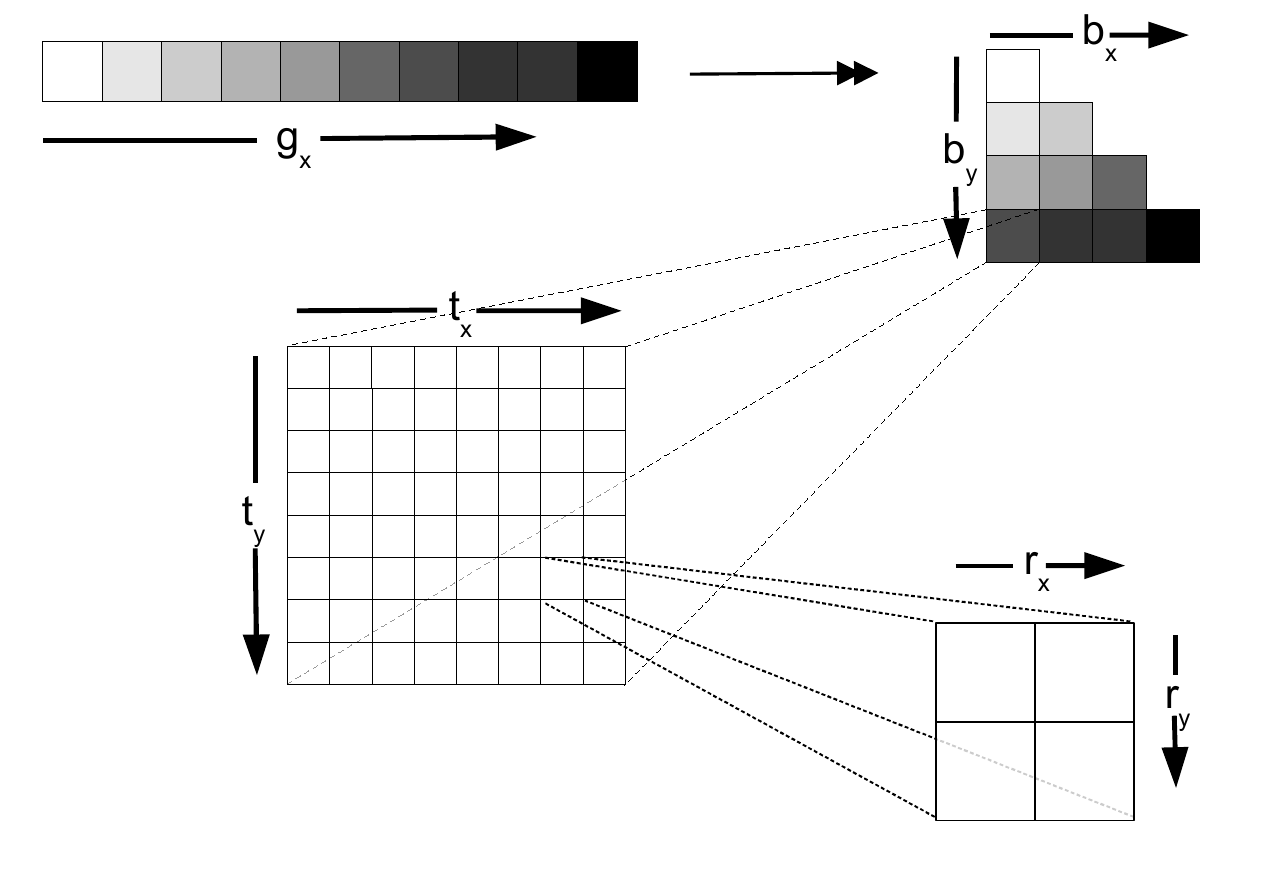}
\end{center}
\caption{Schematic description of the how threads are mapped to the
 correlation matrix.  The linear grid index \(g_x\) is mapped to the
 triangular block index \((b_x,b_y)\).  Each thread \((t_x,t_y)\)
 within the thread block is then responsible for calculating an
 \(R_x\times R_y\) tile of the correlation matrix (indexed by
 \((r_x,r_y\))).  The grid index \(g_y\) maps trivially to the
 frequency dimension (not shown).}
\label{fig:mapping}
\end{figure}

We deployed an alternative strategy where we only launch the minimum
required number of thread blocks, i.e., those which lie within the
lower triangular part of the matrix and calculate the global-baseline
index on the fly by mapping the \(x\)-dimension of the grid index to a
2-dimensional triangular-thread-block index: the correlation matrix is
tiled with squares of size \((R_x T_x\times R_y T_y)\), where
\(\TT=(T_x, T_y)\) is the 2-dimensional thread block size, and
\(\RR=(R_x, R_y)\) is the 2-dimensional register tiling size.  We
impose the constraint \(R_x T_x = R_y T_y = RT\) since this simplifies
the triangular packing.  The grid dimensions are set to
\(\overrightarrow{G} = \left(\frac{N}{2RT}(\frac{N}{RT}+1),
  F\right)\), where \(G_x\) is the number of required thread blocks
per frequency channel and \(G_y\) is trivially mapped to the frequency
dimension.  The mapping from the grid index \(g_x\in[0,G_x)\) to the
2-dimensional triangular block index \(\overrightarrow{b} =
(b_x,b_y)\) is given by
\[
g_x = \frac{b_y(b_y+1)}{2} + b_x \quad\mbox{with}\quad b_y \in
[0,\frac{N}{RT}), \quad b_x \in [0, b_y].
\]
This can be inverted by solving the quadratic equation in \(b_y\)
using integer arithmetic,
\begin{eqnarray*}
b_y & = & \lfloor-\frac{1}{2} + \sqrt{\frac{1}{4} + 2g_x}\rfloor\\
b_x & = & g_x - \frac{b_y(b_y+1)}{2}.
\end{eqnarray*}
This can be evaluated efficiently on the device because of the
presence of the fast square-root intrinsic, and in any case need only
be evaluated once prior to the time integration.  The global-baseline
coordinates are given by \((i,j) = (R_x(b_x T_x + t_x), R_y (b_y T_y +
t_y))\) where \(\overrightarrow{t} = (t_x, t_y)\) is the 2-dimensional
thread block index.  The thread block division strategy is illustrated
in Figure \ref{fig:mapping}.

\subsection{Hardware-Managed Cache Implementation}
\label{sec:hardware-cache}

The introduction of the traditional hardware-managed L1 cache to Fermi
rasies the possibility of not having to use an explicit
software-managed cache, which has been commonplace with CUDA
applications prior to Fermi.  Using a hardware-manged cache is much
easier since there is no need to be concerned with explicit thread
synchronization nor I/O load balancing between the threads (see
\S\ref{sec:software-cache}).  In this implementation each thread
iterates through the time dimension, accumulating the result for an
\(R_x\times R_y\) tile of baselines in the registers, and is
responsible for loading all of the data it requires.  For the thread
blocks which occur on the matrix diagonal, if the global thread index
is located in the super-diagonal the thread exits immediately doing no
work.  Once the accumulation is complete the thread writes its result
to device memory.  In order to achieve optimium bandwidth for
device-memory writes, the matrix elements are reordered into a struct of float4
arrays such that consecutive threads write 16 bytes to contiguous
blocks of device memory, e.g., coalescing is achieved at the
quarter-warp level.  The number of float4 arrays in the struct is
determined by the register tiling, e.g., for \(\RR=(1,1)\) there are eight
numbers thus corresponding to two float4 arrays.

To satisfy the input memory requirements each thread must perform
\(R_x\) and \(R_y\) float4 (=16 bytes) loads for each row and column,
respectively.  Thus the size of the row read in will equal \(16 R_x
T_x\) bytes; this will occur at maximum bandwidth when a.) the row
size is a multiple of the L1 cache line size and b.) the memory
transactions can be broken down into 128-byte requests at the full-,
half- or quarter-warp level.  All subsequent requests from this thread
block for the same row will be fully cached assuming no evictions have
taken place, and thus the L1 cache should automatically enable tiling
in the \(y\) dimension of size \(T_y\).  When reading in the column,
each warp will request \(16R_y (32/T_x)\) bytes where \(32/T_x\) is
the number of rows that a given warp extends over.  Again, these
transactions will occur at full bandwidth only when this is equal to
the cache-line size at the full-, half- or quarter-warp level.  For
almost all the tile sizes explored below the bytes requested per warp
for the column load is less than the L1 cache line size, thus one
would expect a drop in cache efficiency.  However, other warps within
the same thread block will likely request the unused fetched
components increasing the cache efficiency.  Since all requests for a
given column entry will originate from within the same warp, tiling
occurs through an L1 broadcast to the \(T_x\) threads along the
corresponding row, thus the effective tiling in the \(x\) dimension is
\(T_x\).  Finally we note, that we also present results with the L1
cache disabled.  We do so to quantify the improvement due to data
reuse in the L1 cache.  We note however, that while there is no
sharing of data between warps, there is a mechanism for cooperation by
memory transactions satisfied by an L2 broadcast within the warp, thus
the effective bandwidth per thread can be significantly higher than
the bandwidth to the L2 cache.

\begin{table}[htb]
\begin{center}
\begin{tabular}{|l|c|c|c|c|c|c|}\hline
                          &                    &         & \multicolumn{2}{|c|}{GFLOPS} & \multicolumn{2}{|c|}{GB/s}\\ \hline
Kernel  & \(R_x\times R_y\) & \(T_x\times T_y\)& L1  on &  L1 off & L1 on & L1 off\\\hline
1 & \(1\times 1\)      & \(8\times 8\)      & 315    & 322    & 319 & 328\\ \hline 
2 & \(1\times 1\)      & \(16 \times 16\) & 411    & 329    & 423 & 339\\\hline 
3 & \(1 \times 1\)     & \(32 \times 32\) & 421    & 337    & 446 & 358\\\hline 
4 & \(1\times 2\)      & \(16 \times 8\)   & 600    & 447    & 458 & 341\\ \hline 
5 & \(2\times 1\)      & \(8 \times 16\)   & 457    & 201    & 349 & 154\\ \hline 
6 & \(2\times 2\)      & \(8 \times 8\)     & 559    & 354    & 288 & 183\\ \hline 
7 & \(2\times 2\)      & \(16 \times 16\) & 623    & 384    & 331 & 204 \\\hline 
\end{tabular}
\caption{Performance and effective bandwidth achieved using the
  hardware-managed cache, both with and without the L1 cache enabled,
  for the range of register and thread block sizes investigated (48
  KiB L1 cache mode).}
\label{tab:hardware}
\end{center}
\end{table}

In Table \ref{tab:hardware} we present the performance results for the
X-engine kernel both with and without cache enabled.\footnote{When
  designing the kernel implementation, we generally tested performance
  at \(N=512\), \(F=6\), and \(I = 1024\).  All results presented in
  this section correspond to these parameters, with scaling results
  delayed until \S\ref{sec:results}.}  In generating these results we
have applied rudimentary optimizations such as loop unrolling, but we
note that in using the hardware-managed cache, there is little scope
for optimization since we are reliant on the hardware.\footnote{Finer
  grained control of the cache can be obtained at the expense of using
  PTX assembly, an option that has become much easier with CUDA 4.0
  which allows inlined assembly instructions.}  The results for
kernels 1, 2 and 3 demonstrate performance where the register tile
size is held fixed at \(\RR=(1,1)\) and the inter-thread tile size is
increased.  While performance increases with increasing tile size, the
improvement is very small in moving from \(\TT=(16,16)\) to
\(\TT=(32,32)\) suggesting that the bottleneck at this point is not
the device-memory bandwidth.  With the cache disabled, the performance
is essentially flat, showing that the improvement with increasing
thread block size is due to increased data reuse.  The effective
memory bandwidth reported with the cache disabled is significantly
higher than the L2 cache bandwidth, which we conclude is due to L2
broadcasts within a warp.  In the subsequent two kernels, both
rectangular register tiling and rectangular thread block sizes are
used, with kernel 5 corresponding to the transpose of the kernel 4.
Kernel 4 shows that increasing the register tiling improves
performance to 600 GFLOPS; this shows that the L1 cache bandwidth is
the limiting factor in the \(\RR=(1,1)\) kernels once the
device-memory bottleneck has been overcome.  Its transpose, however,
shows negligible improvement over the \(\RR=(1,1)\) kernels; this is
because in reading the row values, each thread now has to read in
consecutive 32 bytes which equates to 256 bytes at the quarter-warp
level.  Such memory transactions are not coalesced, and incur double
the number of memory transactions, vastly decreasing performance.
This reduced bandwidth is most evident with the L1 cache disabled.
Finally we have kernels 7 and 8, which use \(\RR=(2,2)\), where it can
be seen that the performance is comprable with the kernel 4 kernel.
Like kernel 5, the row reads will not be coalesced since the
transaction size is 256 bytes at the quarter-warp level.  Thus any
improvement from increasing the register tiling is offset by a
reduction in device memory bandwidth.

\subsection{Software-Managed Cache Implementations}
\label{sec:software-cache}

We now turn our attention to using the shared memory to enable
inter-thread tiling.  Since all memory is explicitly managed,
programming for the shared memory has the potential for much greater
performance since full memory coalescing can be obtained regardless of
the register tile size used.  The approach can summarized as follows:
all memory load requests are undertaken cooperatively by all threads
in the thread block, where some threads will load the required row,
and others will load the required column.  These numbers are then
communicated through the shared memory before computation proceeds as before.

For a thread block of size \(T_x\times T_y\) and register tiling of
size \(R_x\times R_y\), we require \(4(R_x T_x + R_y T_y)\) real
numbers.  Each number should only be loaded once, and all \(T_xT_y\)
threads should partake in memory transactions to maximize latency
hiding.  To satisfy this constraint we require that
\[
\frac{4(R_x T_x + R_y T_y)}{V} = T_x T_y,
\]
where \(V\) is the number of floats loaded per thread, e.g., \(V=2\)
equates to each thread loading a float2.  This constraint means that
cannot test software-managed cache variants of kernels 2 and
3.\footnote{It would be possible to admit the excluded \(\TT=(16,16)\)
  and \(\TT=(32,32)\), with \(\RR=(1,1)\) parameters if we were
  willing to allow sub-32-bit memory transactions per thread.  Another
  option we did not explore was to allow \(V\) to vary for the row and
  column loads, e.g., load in the row values using floats while
  loading the column values using float2s.}


In this approach the first \(4R_xT_x/V\) threads each load the
required row elements, and the remaining \(4R_yT_y/V = T_xT_y -
4R_xT_x/V\) threads load in the column elements.  Warp divergence does
not take place because warps are never split between row and column
loads.  Although different warps are used to load in row and column
values, warp dependendent branching does not occur in the accumulation
because all memory address pointers for each thread are computed prior
to the accumulation loop, and are then simply iterated by a constant
amount at each iteration of the loop.  Full bandwidth to device memory
on both row and column reads is obtained by employing a different
thread ordering scheme for the memory reads from that used for the
computation, this is facilitated by the shared memory to distribute
the fetched data to the required threads.  When the writes to and the
subsequent reads from shared memory are performed, the access patterns
are chosen to avoid bank conflicts, e.g., adjacent threads store
32-bit words in adjacent memory banks.  When reading the required
numbers back from shared memory into registers the multicast ability
of the Fermi architecture is critical to ensure that full
shared-memory bandwidth is achieved.  As a result, the inter-thread
tile size is given simply by the product \(R_xT_x\times R_y T_y\).

Extra care must be taken for the thread blocks that lie on the matrix
diagonal.  Since all threads cooperate for the element loading, any
threads that occur in the matrix super-diagonal must not exit
prematurely.  We simply let all threads perform all computation, but
only have the threads corresponding to the diagonal and sub-diagonal
actual write their result to device memory.  We return to this issue
of wasted computation along the matrix block diagonal in
\S\ref{sec:kernel-performance}.

\begin{table}[htb]
\begin{center}
\begin{tabular}{|l|c|c|c|c|c|c|c|c|}\hline
 &                           &                             & \multicolumn{3}{|c|}{GFLOPS} & \multicolumn{3}{|c|}{GB/s}\\ \hline
Kernel & \(R_x\times R_y\) & \(T_x\times T_y\)&  Initial & Buf & Tex & Initial & Buf & Tex\\\hline
8 & \(1\times 1\)      & \(8\times 8\)      &   414 & 562   &   510  & 53 & 72 & 65\\ \hline %
9 & \(1\times 2\)      & \(16 \times 8\)   &  771 &  837 &  818  & 49 & 54 & 53\\ \hline %
10 & \(2\times 1\)      & \(8 \times 16\)   &  783 &  852  &  834 & 50 & 55 & 54 \\ \hline %
11 & \(2\times 2\)      & \(8 \times 8\)     &  801  & 944   & 1023  & 53 & 61 & 66 \\ \hline %
12 & \(2\times 2\)      & \(16 \times 16\) & 644  &  908  &  897   & 21 & 31 & 30 \\ \hline %
\end{tabular}
\caption{Performance and bandwidth using the software-managed cache
  (Initial -- initial implementation, Buf -- double buffered shared
  memory, Tex -- memory reads performed using texture
  cache, 48 KiB shared memory mode).}
\label{tab:software}
\end{center}
\end{table}

The performance and sustained memory bandwidth of the initial software
managed cache implementions are given in Table \ref{tab:software}
(under the heading Initial).  When compared to the hardware-managed
cache implementations in Table \ref{tab:hardware}, we see that
performance is alway better using the software-managed cache for given
\(\RR\) and \(\TT\) parameters (i.e., 1 \& 8, 4 \& 9, 5 \& 10, 6 \& 11
and 7 \& 12).  Note that the two rectangular tiling kernels (9 and 10)
have very similar performance unlike the case for the hardware-managed
cache becuase full memory coalescing is now obtained for both
variants.  Perhaps surprisingly, the performance of kernel 12 is the
second slowest, despite it having the greatest inter-thread tile size
and joint highest register tiling.  This can be primarily attributed
to the required thread synchronization between reads from and write to
shared memory. Although all kernel variants require thread
synchronization, because of the larger thread block size, no other
thread blocks can run concurrently on the same SM, thus thread
synchronization causes a large performance stall.

In looking at the achieved bandwidth by all of the software-managed
kernels, we see that less than 50\% of deivce-memory bandwidth is
sustained.  Since all the memory transactions should be fully
coalesced, we conclude that none of the kernels are device-memory
bandwidth bound at these parameters, and are either shared memory
bandwidth bound, latency bound or instruction bound.  Note that this
measure of bandwidth measures actual bandwidth obtained into the
registers and cannot be compared with that reported in Table
\ref{tab:hardware} which measures the effective bandwidth due to the
cache.


Each iteration of the accumulation loop requires two thread
synchronizations because we must ensure that threads a.) do not write
into shared memory before threads still in the previous iteration have
finished reading from shared memory, and b.) do not read from shared
memory until all threads have written their data into shared memory.
We note that while these two conditions cannot be avoided, we can half
the number of thread synchronization points by double buffering the
shared memory: on even iterations of the accumulation loop we will
read from buffer 0, but write to buffer 1, and vice versa for odd
iterations. This double buffering combined with an additional by-hand
loop unrolling by two corresponds to the performance figures presented
in Table \ref{tab:software} entitled ``Buf''.  In making this change,
we see a significant performance improvement for all kernels, but
especially so for kernel 12, which we hypothesized previously was most
bound by thread synchronization.
 
The final performance reported in Table \ref{tab:software}, entitled
``Tex'', is where the all memory reads are performed using the texture
unit.  The rationale here was not to gain benefit using the primitive
cache offered through texture reads, which is generally slower than
the L1 cache, but rather to use the linear interpolation unit to
perform the global array indexing for free, removing the explicit
pointer arithmetic from the kernel.  Making this change generally
proved detrimental, with the important exception of kernel 11, which
increased performance to in excess of 1 TFLOPS.


In this section we have shown that using a multi-level tiling
algorithm together with a software managed cache is critical for
achieving maximum performance.  All subsequent results will
exclusively use kernel 11, since this has the best performance.

\section{Bus transfers}
\label{sec:pcie}


The communication over the bus can be overlapped with kernel execution
using CUDA's asynchronous API.  As long as the total time spent in
communication is less than that spent in computation, the
communication can in principle be hidden allowing the X-engine to
operate at peak.  A subtle point that has not been explicitly
treated in previous work on similar applications is that the total
bandwidth to device memory must be shared between the kernel and the
asynchronous PCIe memory transfers~\cite{Paulius}.  This may be partly
the reason that the performance reported in
\cite{Nieuwpoort09usingmany-core} declined when the overlapping of
kernel execution and bus transfers were employed.  In the present
case, since the kernel is not bandwidth bound, sustaining around 37\%
of peak memory bandwidth to device memory, the expectation is that
there should be ample memory bandwidth to sustain both the kernel and
concurrent PCIe bus transfers at maximum rate.  Note this is distinct
from the question as to whether the achievable PCIe bus transfer rate
is fast enough to meet the data requirements of the kernel.

\begin{algorithm}[htb]
\SetKwInOut{Input}{input}
\SetKwInOut{Output}{output}
\SetKwInOut{Allocate}{allocate}
\Input{input\_h[\(N_p\)][\(I' F N\)] \tcp{input signal vector}}
\Output{matrix\_h[\(\frac{1}{2}F N(N+1)\)] \tcp{packed correlation matrix}}
\BlankLine
\Allocate{buffer\_d[2][\(I' F N\)] \tcp{signal vector buffers}}
\Allocate{matrix\_d[\(\frac{1}{2}F N(N+1)\)] \tcp{packed correlation matrix}}
\BlankLine
\tcp{Execution pipeline}
buffer\_d[0] \(\leftarrow\) input\_h[0]\;
\For{\(p\leftarrow 1\) \KwTo \(N_p\)}{
matrix\_d \(\leftarrow\) X-engine(matrix\_d, buffer\_d[\((p+1)\bmod{2}\)])\;
buffer\_d[\(p\bmod{2}\)] \(\leftarrow\) input\_h[p]\;
Synchronization\;
}
matrix\_d \(\leftarrow\) X-engine(matrix\_d, buffer\_d[\((N_p+1)\bmod{2}\)])\;
\BlankLine
\tcp{Transfer result back to host}
matrix\_h \(\leftarrow\) matrix\_d\; 
\label{alg:pipeline}
\caption{Integrated X-engine communications pipeline pseudocode.  The
  suffixes \_h and \_d denotes memory buffers allocated on the host
  and device respectively.  For convenience we adopt a two-dimensional
array notation for the input signal vectors.}
\end{algorithm}

To allow overlapping of communicaton and computation, these operations
must operate on independent data.  The X-engine execution consists of
a three stage pipeline: 1.) host\(\rightarrow\)device transfer of
data, 2.)  kernel execution, and 3.)  device\(\rightarrow\)host
transfer of the final correlation matrix.  For a large enough
integration length \(I\) the final transfer back to the host is a
negligible contribution to the total time.  Additionally we expect for
many applications, it would be desirable to retain the correlation
matrix on the device for further post-processing, e.g., for real-time
imaging and callibration~\cite{Edgar:2010uq} or scrubbing of man-made
radio interference~\cite{kocz2010}.  We thus aim to overlap only the
host\(\rightarrow\)device transfer with the kernel execution.  To this
end, we break the input sample vector of length \(I F N\) into \(N_p\)
sample vectors of length \(I' F N\) with \(I = N_p I'\).  We then loop
over the \(N_p\) sample vectors, at each iteration calculating the
outer-product sum and summing the result to the correlation matrix.
This allows the overlap of communication and computation, since the
\((p+1)^{th}\) sample vector can be transferred to the device while
the \(p^{th}\) sample vector's contribution to the correlation matrix
is computed.  At each iteration, explicit synchronization is
enforced to prevent a race condition between the memory transfer and
kernel execution.  The execution pipeline is illustrated in Algorithm
\ref{alg:pipeline}.

\begin{figure}[htb]
\begin{center}
\includegraphics[width=10cm]{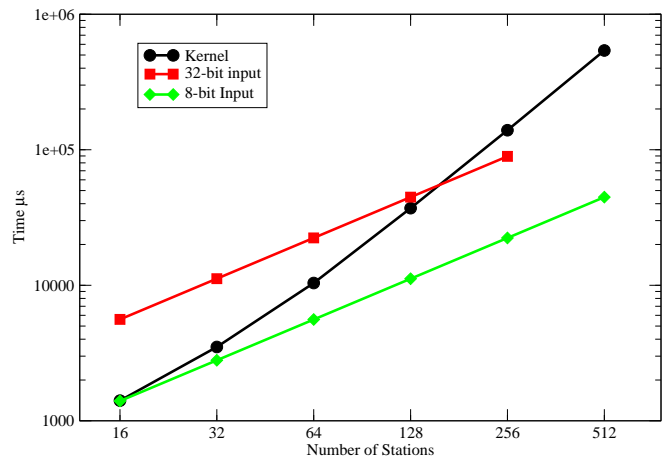}
\end{center}
\caption{Kernel-only execution time, and host\(\rightarrow\)device
  transfer time for 32-bit and 8-bit data as a function of number of
  stations (\(F=128, I = 1024\)).  With 32-bit precision, there is not
  enough device memory to accomodate \(N=512\) at these parameters.}
\label{fig:PCIe}
\end{figure}

Figure \ref{fig:PCIe} is a plot comparing the execution time of the
X-engine with the host\(\rightarrow\)device transfer time over the
PCIe bus as a function of number of stations.  (We defer the
discussion of how actual performance varies as the number of stations
until \S\ref{sec:results}.)  For a large number of stations
(\(N>128\)), the quadratic scaling of the X-engine ensures that the
communication can be completely overlapped with computation, and the
kernel can operate at peak performance.  However, at small \(N\) the
host\(\rightarrow\)device communication time dominates the computation
time resulting in severely impaired performance.  This plot agrees
with the prediction made in \S\ref{sec:mapping} regarding the minimum
tile size \(N=105\) required to overcome the PCIe bottleneck.  Without
data flow reduction, the X-engine would appear to be severely bound by the
PCIe bandwidth.

However, use of 32-bit floating point data is unnecessary for the
X-engine.  The precision of data input to the X-engine typically
depends on the incidence of impulsive manmade interference.  Strong
interference militates for more bits (e.g., 12 bits can represent a
dynamic range of \(\sim 10^7\)), but truncation to 8 bits is usually
safe (\S\ref{sec:fpga}).  The kernel is easily adapted to accomodate
this using the texture read mode {\it cudaReadModeNormalizedFloat}:
8-bit data is read through the texture cache, and is converted to the
native 32-bit floating point format using the texturing hardware,
i.e., without subtracting the number of cycles available for
computation.  Since the kernel is not device-memory bandwidth bound,
this change has no effect upon the raw-kernel
performance.\footnote{Such reduction in memory bandwidth requirements
  does potentially improve the scalability of the current algorithm
  (i.e., two levels of tiling) to future possible architectures where
  the required arithmetic intensities to achieve near peak peformance
  would be expected to be considerably higher than current GPUs.}
However, it reduces by a factor of four the transfer time over the
PCIe bus for the input data.  The result is that communication can be
hidden for \(N > 16\) stations.  A desirable side effect is the
reduction in memory footprint allowing many more frequency channels or
a longer time integration to be performed.  Note the ratio of the
kernel execution time to the host\(\rightarrow\)device transfer time
is independent of both the number of frequency channels and the
integration length and so this conclusion is universal (this is only
approximately true -- see \S\ref{sec:results}).

\section{Performance Results}
\label{sec:results}

As stated previously all performance results utilize kernel 11, and
here we additionally restrict ourselves to 8-bit input data and we shall
only consider cases where the the number of stations is a power of
two.

\subsection{Kernel Performance}
\label{sec:kernel-performance} 
\begin{figure}[htb]
\begin{center}
\includegraphics[width=10cm]{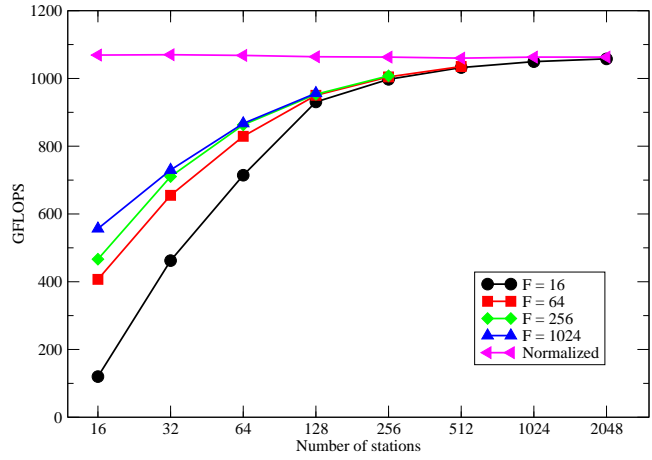}
\end{center}
\caption{Performance of the X-engine as a function of the number of
  stations (\(N\)) and frequency channels (\(F\)) and fixed
  integration length \(I=1024\).  Memory limitations prevent large
  \(F\) and \(N\) simultaneously.  The magenta curve is the normalized
  GFLOPS including the wasted compution performed on the matrix
  diagonal.  The slight decline in performance of the normalized
  values with increasing \(N\) is likely due to the quadratically
  increasing output memory traffic not being completely supressed at
  this value of \(I\).}
\label{fig:station}
\end{figure}

\vspace{2mm}
\paragraph{Performance versus Number of Stations}
Performance increases significantly as we increase \(N\) (at fixed
\(F\)), plateauing to a peak of 1058 GFLOPS, which equates to 79\% of
the theoretical peak performance of the GTX 480 (Figure
\ref{fig:station}).  When we include the additional overhead of
required auxillary instructions; e.g., load/store, thread
synchronization; we find the kernel is actually operating at 91\% of
peak performance.  To our knowledge, no other non-trivial application
reaches this level of performance.  At small \(N\) performance is
impaired because: 1.) the correlation matrix is smaller, hence fewer
threads are active; and 2.) there is a proportional increase in the
amount of wasted computation perfomed by the thread blocks distributed
along the matrix diagonal.  The first of these effects can be offset
by increasing the number of frequency channels to increase the number
of thread blocks.  This is demonstrated on Figure~\ref{fig:station}
where we see that increasing the number of frequency channels
drastically improves performance at low \(N\), while having a
negligible effect at large \(N\).

Beyond this, reduced performance at small \(N\) results from the
wasted computation performed on the threads blocks which appear along
the diagonal of the correlation matrix.  The total or ``normalized''
performance is the sum of the desired and wasted computation.  Here we
use the largest \(F\) available at a given \(N\) to eliminate the
effect of low thread occupancy.  There are \(N/RT\) thread blocks
along the matrix diagonal, with each thread block corresponding to
\(RT\times RT\) matrix elements.  Thus, the additional wasted work
scales as \(\sim(N/RT) 2RT(2RT-1)\) (i.e., the strictly upper diagonal
of the diagonal thread blocks).  Since the total desired computation
scales as \(\sim 2N(2N+1)\), the total computation rate including the
wasted operations is found from multiplying by \(\frac{2(N+RT)} {2N +
  1}\).\footnote{This correction factor is only exact in the limit of
  infinite arithmetic intensity because it is only unecessary floating
  point operations that are performed, no additional memory traffic is
  incurred.  Thus at finite, but large, arithmetic intensity we expect
  the correction factor to be a small overcorrection.}  Using this
normalized metric the performance is independent on the number of
stations.  For \(N < 256\) the difference between actual and the
normalized performance is appreciable, suggesting that a kernel
specfically optimized for the diagonal thread blocks at small \(N\)
may be worth considering.

\paragraph{Integration length versus Number of Frequency Channels}
As described in \S\ref{sec:character} the effect of the integration
length \(I\) upon performance is expected to become significant at
small \(I\) since the output memory traffic is supressed as
\(I^{-1}\).  We illustrate this in Figure \ref{fig:int_time} where we
vary \(F\) and \(I\) subject to keeping the product \(F I\) constant.
This corresponds to a fixed amount of computation per input,
equivalent to Equation \ref{eq:flops} at fixed bandwdith \(B\).  The
increase in \(F\) necessarily results in a shortening of the
integration length, thus at some point the resulting increase in
memory traffic reduces performance.  Where output memory traffic is
ignored we observe, owing to data parallelism, expected high
performance without roll off as \(F\) is increased (\(I\) is
decreased).  The difference is most evident at small \(N\) since many
more frequency channels are achievable.

Obtaining high performance can thus be a balancing act between
increasing the concurrency for computation against the resulting
increase in memory bandwidth requirements.  The parameter space of our
demonstration is limited by the parameters of the GTX 480.  While this
is not a problem for that platform at large \(N\) and supportable
ranges of \(F\), GPUs with larger memory than the GTX 480 and greater
required parallelism, i.e., more cores, will exhibit pronounced roll
off for larger \(F\) (smaller \(I\)) at these values of \(N\).  

Even for a fully optimized X-engine kernel, there is still further
parameter tuning that must take place in order to obtain maximum
performance.  The locus in parameter space that results in near
maximum performance would be one of many prominent factors in
specification and basic system engineering of a radio telescope.

\subsection{Integrated Performance}

\begin{figure}[htb]
\begin{center}
\includegraphics[width=10cm]{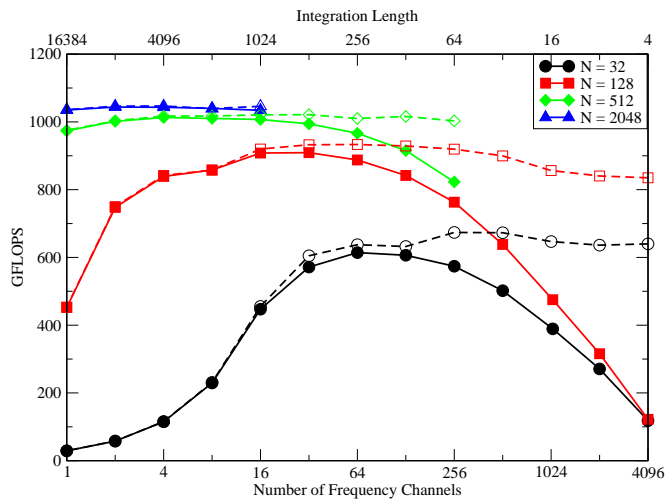}
\end{center}
\caption{Performance of the X-engine kernel for a fixed volume of data
  (16384) per input and different numbers of stations (\(N\)).  The
  solid curves indicate actual X-engine performance, whereas the
  dashed curves indicate performance with the output of the
  correlation matrix disabled.  Memory limitations prevent the use of
  a shorter integration length for \(N=\) 512 and 2048 curves since
  this increases \(F\) and with it the number of correlation matrices
  to be output.  In general as we increase the volume of data to be
  processed (i.e., \(FI\)), the initial increase in performance with
  increasing \(F\) will remain fixed, but the performance plateau and
  subsequent drop in performance will stretch since \(I^{-1}\) memory
  output supression and \(F\) enhancement will cancel each other out.}
\label{fig:int_time}
\end{figure}

\begin{figure}[htb]
\begin{center}
\includegraphics[width=10cm]{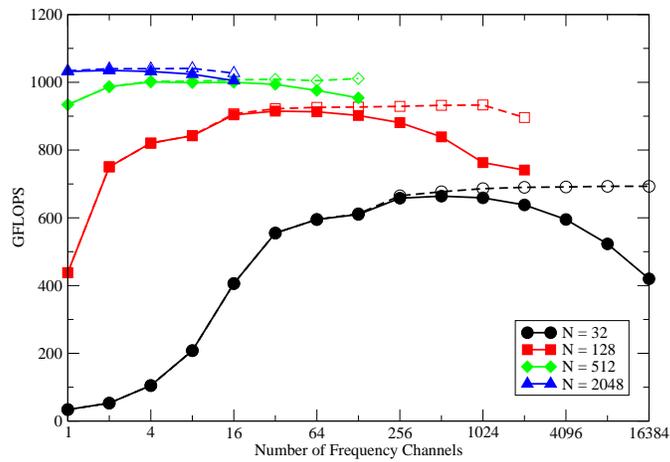}
\end{center}
\caption{Performance of the integrated X-engine as a function of
  number of frequency channels and number of stations (\(N\)) for a
  fixed 1 GiB of 8-bit input data.  The GPU integration length
  \(I'=256,1024,2048,2048\) for \(N=32,128,512,2048\), respectively.
  The solid curves indicate performance when the correlation matrix is
  transferred to the host, whereas the dashed curves are for when it
  is kept on the device.}
\label{fig:integrated}
\end{figure}

The performance of the integrated X-engine is not affected by
host\(\rightarrow\)device transfer overhead for a fixed input of 1
GiB, which corresponds to around 0.2 seconds of PCIe transfer time
(Figure \ref{fig:integrated}).  As described in Algorithm
\ref{alg:pipeline}, the input vector of length \(I F N\) is
split into \(N_p\) vectors of length \(I'FN\), where we chose the
kernel integration length \(I'\) to maximize performance based on the
results obtained in \S\ref{sec:kernel-performance}, i.e., we chose
\(I'=256, 1024, 2048, 2048\) for \(N=32,128,512,2048\), respectively.
In fact performance is actually higher here at \(N=32\) than in
Figure~\ref{fig:int_time} because the vector of data is larger, i.e.,
with a larger number of frequencies and a longer integration length.
Only for short integration lengths, i.e., a large number of frequency
channels, does the overhead of the device\(\rightarrow\)host transfer
become apparent.  The overall shape of the curves in Figures
\ref{fig:int_time} and \ref{fig:integrated} are similar because both
convey the impact of bandwidth constraints (albeit from different
source) for output of the correlation matrix on performance.

\begin{figure}[htb]
\begin{center}
\includegraphics[width=10cm]{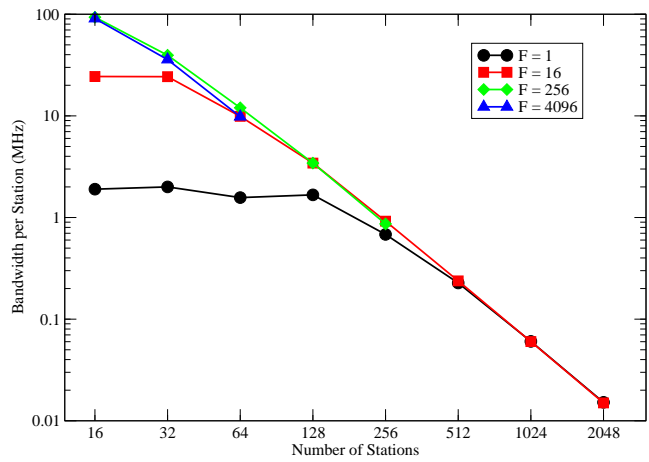}
\end{center}
\caption{Bandwidth per station (assuming dual polarization) for the
  integrated X-engine as a function of number of stations for \(F = 1,
  16, 256, 4096\) with 1 GiB of 8-bit input data.  The curves indicate
  achievable bandwidth including all PCIe bus transfers.}
\label{fig:bandwidth_station}
\end{figure}

To interpret performance figures in the context of radio telescope
system design, we have estimated the achievable correlated
dual-polarization signal bandwidth as a function of the number of the
stations (Figure \ref{fig:bandwidth_station}).  This can be used to
estimate the number of GPUs required to process a total signal
bandwidth given a number of stations, assuming that data for
difference frequency channels can be trivially distributed among
different GPUs in the context of a cluster.  For example, in the case
of a 128-station MWA configuration with 30.72 MHz instantaneous
bandpass the X-engine processes 3.5 MHz of signal bandwidth, thus
requiring 9 GPUs (GTX 480s).

\section{Discussion}
\label{sec:discuss}

\subsection{Comparison against other architectures}
\label{sec:compare}

In comparison to previous GPU implementations of the
cross-correlator~\cite{Harris:2008, Wayth:2009,
  Nieuwpoort09usingmany-core}, we have improved upon the performance
by at least a factor of four.  Half of this improvement can be
attributed to the Fermi architecture, with the rest attributed to the
multi-level tiling strategy and other fine tuned optimizations.
A comparison with the wide range of platforms
presented in \cite{Nieuwpoort09usingmany-core}, reveals that the
GeForce GTX 480 running our implementation is a significant
improvement over all other commodity processors, both in terms of
absolute performance and performance per watt.  This statement stands
even in factoring improvements due to Moore's law on other platforms
since this prior work was published.

A more subtle comparison is required in consideration of FPGAs.  A
direct comparison against current FPGAs for cross-correlation is
difficult, and one must really consider the integrated system design
and engineering for a specific instrument.  We take for example the
correlator for a 128-station MWA, whose detailed specifications are
described in \cite{skamp} in the context of dual-use application to
the SKAMP telescope.  This uses a Virtex 4 FPGA architecture, which
can process the required 30.72 MHz MWA bandwidth per signal for a
power budget of 0.75 kW (X-engine only).\footnote{Here we have
  estimated the power consumption for a 128-station MWA by scaling the
  power consumption from the 512-station MWA
  correlator~\cite{Lonsdale:2009cb}.} In comparison, if we were to
deploy a GTX 480 based X-engine, we have at a minimum 250 W \(\times\)
9 = 2.25 kW neglecting the overhead of the host system power, which
further favors FPGAs.  When one factors in that this FPGA correlator
represents at least two-generations-old technology relative to the GTX
480,\footnote{Fabrication process technology shrinks by roughly 70\%
  every two--three years.  Virtex 4 FPGAs were fabricated at 90 nm and
  all Fermi GPUs are fabricated at 40 nm.} we see that GPUs struggle
to compete based on only the power metric.  However, there may be
longer-term factors that differ, such non-recurring costs, in
development time and procurement.  Radio telescope design and
construction can extend over years, and engineering risk management
may motivate staying at one generation behind the FPGA development
curve.  As well, unit costs for newly introduced FPGAs are high, and
successfully developing optimized, robust bit code requires
specialized engineering training and experience.  In contrast,
maintaining parity with the bleeding edge is more readily achievable
with GPUs, where hardware unit costs are lower, hardware is intended
to be commodity driven, programming leverages high-level environments
(CUDA and OpenCL) that abstract much of the hardware architecture and
are individually forward-backward compatible (release to release and
generation to generation), and successful coding has a strong link to
relatively generic parallel programming practices.  Indeed, since this
research was undertaken, Nvidia has launched a successor card, the
GeForce GTX 580 which is 20\% more power efficient in executing the
X-engine.  The Kepler series due at the end of 2011, and subsequent
generations are expected to improve upon this metric
significantly~\cite{huang}.  The X-engine will run efficiently on
these architectures because of software compatibility, and more
importantly {\it algorithm scalability}.  Thus, an open question
remains, that being whether GPUs can surplant FPGAs for
cross-correlation, much as GPUs have replaced GRAPE for N-body
computations~\cite{Schive:2007mn}.  In this case, although the
fixed-function GRAPE is more power efficient for the problem,
economies of scale and high peak efficiencies obtained using GPUs
essentially have made GRAPE redundent.

\subsection{Future Machines}
Although we have shown that current generation GPUs are extremely
competitive for cross-correlation, the question that arises is whether
this will hold in the future.  Future GPUs will likely continue to
have an exponentially increasing number of cores, with only a linear
increase in memory bandwidth.  The prototypical exascale machines that
are currently envisaged feature a very deep memory hierarchy;
obtaining high performance on such machines will require increasing
the algorithmic arithmetic intensity as one moves away from the
processing cores.  For cross-correlation, the increased arithmetic
intensity requirements can be met by employing an increased number of
tiling levels.  Thus, cross-correlation is likely to be one of a
handful of real scientific applications of interest that will be able
to harness fully at least early generation exascale machines.

\section{Conclusion}
\label{sec:conclusions}
We have presented an implementation of the cross-correlation algorithm
using Nvidia's Fermi architecture.  This application is very well
matched to the architecture, and sustains in excess of 1 TFLOPS for a
large number of stations (79\% of theoretical peak).  Key to obtaining
this performance was the use of a software-managed cache and a
multi-level tiling strategy.  This performance can be sustained when
streaming the input data over the PCIe bus.  This represents a
significant improvement over previous work on GPUs and comparison to
other commodity platforms.  While not power competitive compared to
FPGA solutions, its increased flexibility and lower development costs
make it an attractive solution.

With a fully optimized GPU X-engine implemented, the next step is to
develop this into a full FX correlator.  The design we are pursuing is
a hybrid approach, utilizing an FPGA solution for the F-engine and
likely a packetized switch approach for the corner
turn~\cite{2008PASP..120.1207P}.

Beyond this, future work will include increasing the efficiency of the
X-engine in the small \(N\) regime.  Although the work here primarily
targetted the large \(N\) regime for which there are significant
science drivers~\cite{Lonsdale:2009cb}, most current radio telescope
array installations are of the small \(N\) type.  By maximizing the
efficiency at low \(N\), e.g., 32 stations, this would increase the
applicability of GPU computing for signal correlation today, giving a
viable alternative to the usual FPGA solutions.

\section{Acknowledgements}

We would like to thank Nvidia for support via the Harvard CUDA Center
of Excellence.  This work was supported in part by the National
Science Foundation grant PHY-0835713.  We would like to thank
D. Mitchell, S. Ord, and R. Wayth for helpful discussions.  We thank
Kathryn Hollar and the Harvard SEAS REU program for enabling the
research that contributed to this work.

\clearpage

\clearpage


\bibliographystyle{chicagoa}
\bibliography{correlator}







\end{document}